\documentclass[aps,prl,twocolumn,showpacs,preprintnumbers,amsmath,amssymb,superscriptaddress]{revtex4}%
\usepackage{graphicx}%
\usepackage{dcolumn}
\usepackage{amsmath}
\usepackage{color}
\usepackage{multirow}

\begin{document}


\title{Low superfluid density and possible multi-gap superconductivity in BiS$_2$-based layered superconductor Bi$_4$O$_4$S$_3$}

\author{P.~K.~Biswas}
\email[Corresponding author: ]{pabitra.biswas@psi.ch}
\affiliation{Laboratory for Muon Spin Spectroscopy, Paul Scherrer Institute, CH-5232 Villigen PSI, Switzerland}

\author{A.~Amato}
\affiliation{Laboratory for Muon Spin Spectroscopy, Paul Scherrer Institute, CH-5232 Villigen PSI, Switzerland}
\author{C.~Baines}
\affiliation{Laboratory for Muon Spin Spectroscopy, Paul Scherrer Institute, CH-5232 Villigen PSI, Switzerland}
\author{R.~Khasanov}
\affiliation{Laboratory for Muon Spin Spectroscopy, Paul Scherrer Institute, CH-5232 Villigen PSI, Switzerland}
\author{H.~Luetkens}
\affiliation{Laboratory for Muon Spin Spectroscopy, Paul Scherrer Institute, CH-5232 Villigen PSI, Switzerland}
\author{Hechang~Lei}
\thanks{Present address: Frontier Research Center, Tokyo Institute of Technology, 4259 Nagatsuta, Midori, Yokohama 226-8503, Japan.}
\affiliation{Condensed Matter Physics and Materials Science Department, Brookhaven National Laboratory, Upton, New York 11973-5000, USA}
\author{C.~Petrovic}
\affiliation{Condensed Matter Physics and Materials Science Department, Brookhaven National Laboratory, Upton, New York 11973-5000, USA}
\author{E.~Morenzoni}
\email[Corresponding author: ]{elvezio.morenzoni@psi.ch}
\affiliation{Laboratory for Muon Spin Spectroscopy, Paul Scherrer Institute, CH-5232 Villigen PSI, Switzerland}

\begin{abstract}
The magnetic penetration depth $\lambda$ as a function of temperature in Bi$_4$O$_4$S$_3$ was studied by muon-spin-spectroscopy measurements. The superfluid density of Bi$_4$O$_4$S$_3$ is found to be very low. The dependence of $\lambda^{-2}$ on temperature possibly suggests the existence of two $s$-wave type energy gaps with the zero-temperature values of 0.93(3)~and~0.09(4)~meV. The upturn in the temperature dependence of the upper critical field close to $T_{\rm c}$ further supports multi-gap superconductivity in Bi$_4$O$_4$S$_3$. The presence of two superconducting energy gaps is consistent with theoretical and other experimental studies. However, a single-gap $s$-wave model fit with a gap of 0.88(2)~meV can not be ruled out completely. The value of $\lambda(T)$ at $T=0$~K is estimated to be $\lambda(0)=861(17)$ nm, one of the largest of all known layered superconductors, reflecting a very low superfluid density.
\end{abstract}

\maketitle


Recent years have witnessed a growing interest in superconductivity within layered crystal structure compounds. The recent discovery of superconductivity in the BiS$_2$-based compound Bi$_4$O$_4$S$_3$~\cite{Mizuguchi1,Singh} with a transition temperature ($T_{\rm c}$) of $\approx4.5$~K has further fostered the research interest in layered superconductors. Exotic superconductivity with higher $T_{\rm c}$ and/or with unconventional pairing mechanism has often been found in materials with a layered crystal structure. In this context, the most familiar examples are the high-$T_{\rm c}$ cuprates and the Fe-based superconductors, where the corresponding CuO$_2$ or Fe$_2$\textit{An}$_2$ (\textit{An} = P, As, Se, Te) layer plays an important role in determining the superconducting properties \cite{Bednorz,Kamihara,Hsu}. Similarly, the BiS$_2$ layer is believed to be the basic building block for inducing superconductivity in Bi$_4$O$_4$S$_3$ and it is expected that the doping mechanism resembles that of cuprates and Fe-based superconductors. This argument is supported by the discovery of other BiS$_2$-based superconductors ReO$_{1-x}$F$_x$BiS$_2$ (Re = La, Ce, Pr, and Nd) with $T_{\rm c}$ values of 10.6, 3.0, 5.5, and 5.6~K, respectively \cite{Mizuguchi2,Xing,Jha,Demura}.

Up to now, only very few theoretical and experimental studies have been performed on Bi$_4$O$_4$S$_3$. The parent phase (Bi$_6$O$_8$S$_5$) is a band insulator which becomes superconducting after electron doping~\cite{Mizuguchi1}. A first principles band structure calculation indicates that the superconductivity in Bi$_4$O$_4$S$_3$ originates from two Bi 6$p$ orbitals, as they intersect the Fermi surface~\cite{Usui}. A RPA analysis of a two-orbital model for the BiS$_2$-based superconductors suggests that the spin fluctuations promote competing $A_{\rm 1g}$ and $B_{\rm 2g}$ superconducting states with similar pairing strengths~\cite{Martins}. Transport measurements under pressure reveal that $T_{\rm c}$ of Bi$_4$O$_4$S$_3$ decreases monotonically without distinct change of the normal state metallic behavior~\cite{Kotegawa}.
The exact nature of the superconducting state is still debated. Hall, Seebeck coefficients, and magnetoresistance measurements suggest exotic multi-band features in Bi$_4$O$_4$S$_3$~\cite{Tan}. Recent measurements of the temperature dependence of the magnetic penetration depth $\lambda$ using the tunnel diode oscillator technique~\cite{Shruti} indicate a conventional \textit{s}-wave type superconductivity in Bi$_4$O$_4$S$_3$. However, the tunnel diode data were collected only down to 1.6 K and hence remain elusive in determining the true symmetry of the superconducting order parameter in Bi$_4$O$_4$S$_3$. Other experiments reported very strong coupling superconductivity with $\Delta_{\rm 1}(0)/k_{\rm B}T_{\rm c} \approx 16.6$ and with a large region of superconducting fluctuations above $T_{\rm c}$ surviving at large magnetic fields
\cite{Li} or impurity driven superconductivity \cite{Sathish}
In this respect it is important to characterize the superconducting and magnetic properties of Bi$_4$O$_4$S$_3$ at a microscopic level and to obtain a detailed knowledge of the superconducting order parameter over the full temperature range.

In this Rapid Communication, we report on the results of muon-spin rotation and relaxation ($\mu$SR) studies of the magnetic penetration depth as a function of temperature and on the symmetry of the superconducting gap in the Bi$_4$O$_4$S$_3$ superconductor. Our results show that the superfluid density in Bi$_4$O$_4$S$_3$ is very low. The observed $\lambda^{-2}(T)$ is found to be well described by a two-gap $s$-wave model, as seen in many layered Fe-based superconductors. A single \textit{s}-wave gap model is also able to give a satisfactory fit of the data, albeit with lower statistical significance. For the two-gap model, the gap values at absolute zero are $\Delta_{\rm 1}(0)=0.93(3)$, $\Delta_{\rm 2}(0)=0.09(4)$~meV, respectively. The corresponding gap to $T_{\rm c}$ ratios are $\Delta_{\rm 1}(0)/k_{\rm B}T_{\rm c}=2.38(7)$ and $\Delta_{\rm 2}(0)/k_{\rm B}T_{\rm c}=0.22(9)$. These values are close to those obtained for various Fe-based superconductors~(see Ref.~\onlinecite{Biswas3} and references therein). We obtain the penetration depth, $\lambda(0)=861(17)$~nm.


Polycrystalline samples of Bi$_4$O$_4$S$_3$ were prepared by a solid state reaction method~\cite{Mizuguchi1}. Sample characterization measurements were performed at the Brookhaven National Laboratory, USA. Transport measurements at different applied fields were performed at the Paul Scherrer Institute (PSI), Villigen, Switzerland. Zero-field (ZF) and transverse-field (TF) $\mu$SR experiments were carried out on the DOLLY and LTF $\mu$SR instruments, located respectively at the $\pi$E1 and $\pi$M3 beamlines at PSI. The sample in the form of a pressed pellet was mounted on a copper fork-shaped sample holder in the DOLLY and on a silver plate in the LTF instrument. The sample was cooled from above $T_{\rm c}$ to base temperature at $H=0$ during ZF-$\mu$SR measurements and in a field of 300~Oe in TF-$\mu$SR measurements. The typical counting statistics were $\sim20$ million muon decays per data point. The ZF- and TF-$\mu$SR data were analyzed using the free software package MUSRFIT~\cite{Suter}.


\begin{figure}[htb]
\includegraphics[width=1.0\linewidth]{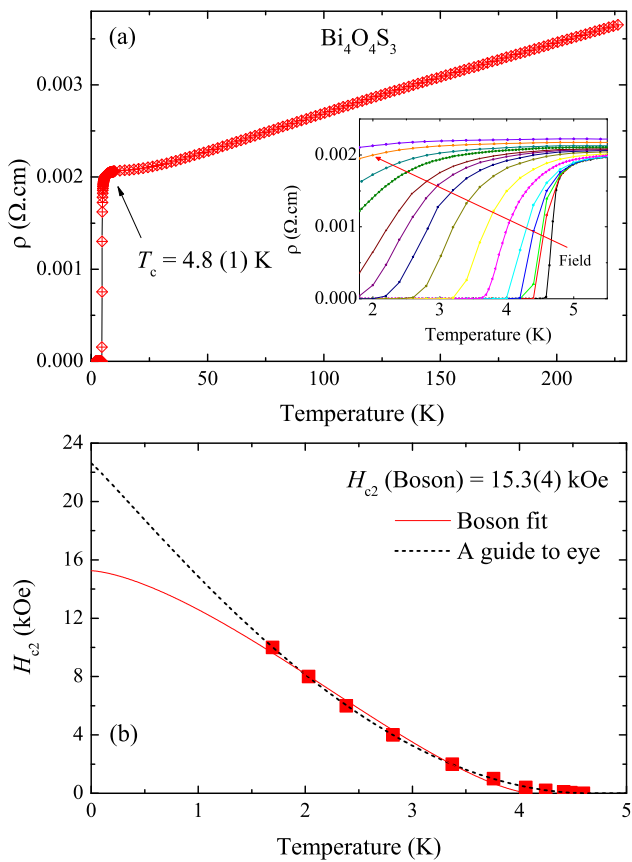}
\caption{(Color online) (a) Temperature dependence of the ac resistivity of Bi$_4$O$_4$S$_3$. Inset shows the temperature variation of the resistivity in a set of magnetic fields of 0, 0.05, 0.1, 0.2, 0.4, 1, 2, 4, 6, 8, 10, 15, 20, 30, and 40~kOe. (b) Temperature dependence of the upper critical field of Bi$_4$O$_4$S$_3$. The solid line is a fit to the data using an expression for charged boson. The dotted line is an empirical interpolation, which better reproduces the temperature region close to $T_{\rm c}$.}
 \label{fig:BiOS_Hc2_v1}
\end{figure}

The ac resistivity of Bi$_4$O$_4$S$_3$ was measured as a function of temperature using a standard four-probe method in a Physical Property Measurement System (PPMS). Figure~\ref{fig:BiOS_Hc2_v1}~(a) shows the temperature dependence of the resistivity $\rho(T)$. A metallic like behavior is observed between 5~K to 220~K with a relatively high residual resistivity value of 2~m$\Omega$.cm. The transition to the superconducting state is relatively sharp with a $T_{\rm c}$~(onset) of 4.8(1) K and $T_{\rm c}$~(zero) of 4.6(1) K. The inset shows the temperature variation of the resistivity in a set of magnetic fields from 0 to 40~kOe. $\rho(T)$ exhibits positive magnetoresistance (MR) in the normal state (just above $T_{\rm c}$). At 6~K, the value of MR increases by $10.6~\%$ for an applied magnetic field of 40~kOe. The temperature dependence of the upper critical field $H_{\rm c2}$ shown in Fig.~\ref{fig:BiOS_Hc2_v1}~(b), was determined from the resistive transitions, defined by a $90~\%$ drop of the normal state resistivity values, determined just above $T_{\rm c}$. The $H_{\rm c2}(T)$ curve shows a positive curvature close to $T_{\rm c}$ and is linear thereafter. A similar behavior has also been observed in polycrystalline borocarbides~\cite{Shulga}, MgB$_2$~\cite{Shigeta,Takano}, Nb$_{0.18}$Re$_{0.82}$~\cite{Karki}, and Re$_3$W~\cite{Biswas} and is considered as a signature of multi-gap superconductivity. A conventional Werthamer-Helfand-Hohenberg model can not describe $H_{\rm c2}(T)$ of Bi$_4$O$_4$S$_3$. However, to estimate $H_{\rm c2}$ at absolute zero, a fit to the $H_{\rm c2}(T)$ data was made using the $H_{\rm c2}(T)$ expression for charged bosons~\cite{Micnas},
\begin{equation}
H_{\rm c2}(T)=H_{\rm c2}(0)\left\{1-\left(\frac{T}{T_{\rm c}}\right)^{3/2}\right\}^{3/2},
 \label{eq:Boson_Hc2}
\end{equation}
The fit yields $H_{\rm c2}(0)=15.3(4)$~kOe, which implies a Ginzburg coherence length $\xi \approx 12$~nm. To get a better estimation of $H_{\rm c2}(T)$ close to $T_{\rm c}$, an empirical interpolation (shown as a dotted line in Fig.~\ref{fig:BiOS_Hc2_v1}~(b)) was later used in Eq.~\ref{eq:Brandt_equation1} for calculating the temperature dependence of the penetration depth of Bi$_4$O$_4$S$_3$.

\begin{figure}[htb]
\includegraphics[width=1.0\linewidth]{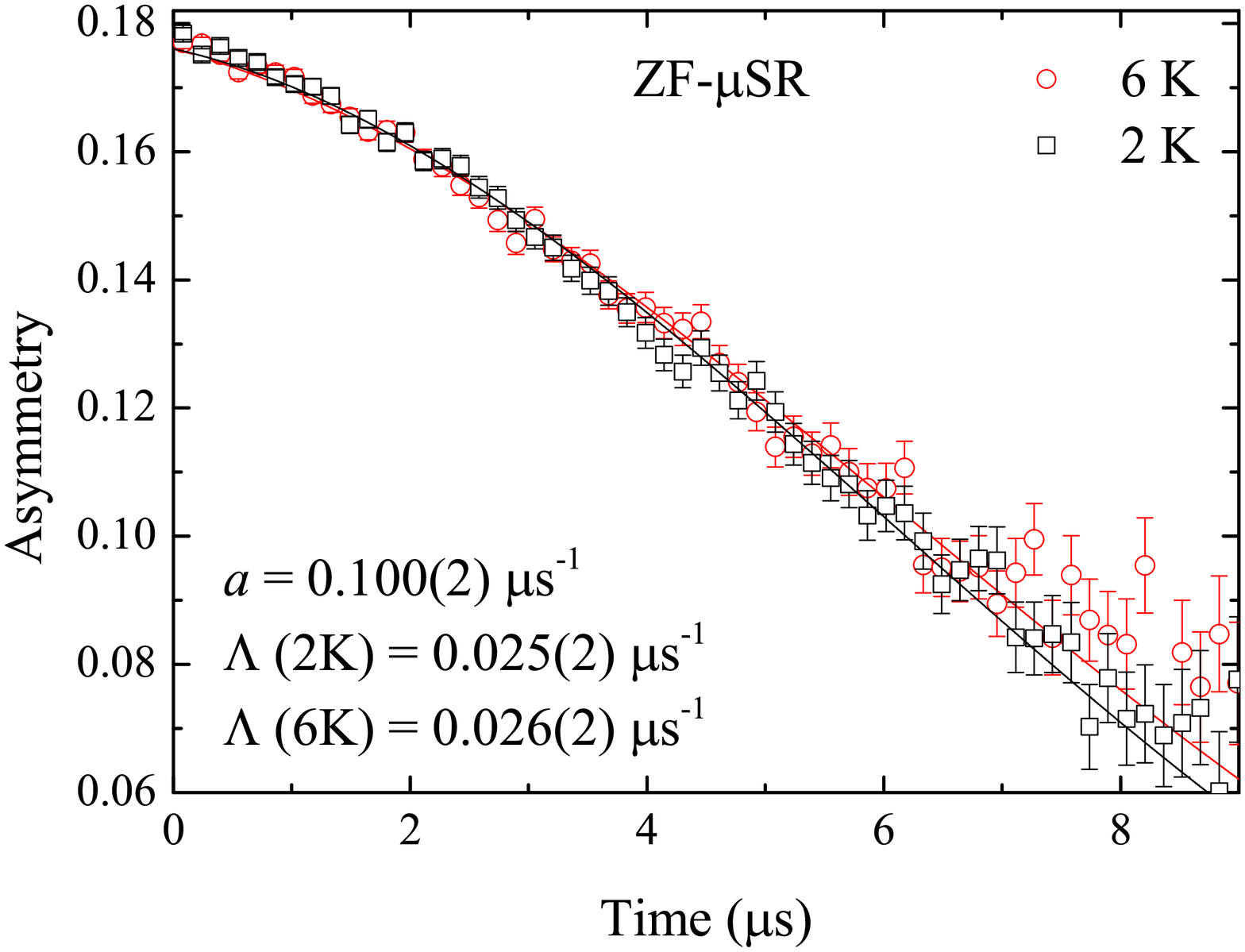}
\caption{(Color online) ZF-$\mu$SR time spectra collected at 2~K and 6~K for Bi$_4$O$_4$S$_3$. The solid lines are the fits to the data using the Eq.~\ref{eq:KT_ZFequation}, described in the text.}
 \label{fig:BiOS_AsyZF}
\end{figure}

ZF-$\mu$SR measurements were performed to look for any unusual temperature-dependent relaxation processes, which may be associated with the onset of superconductivity. Figure~\ref{fig:BiOS_AsyZF} shows the ZF-$\mu$SR time spectra of Bi$_4$O$_4$S$_3$, collected below (at 2~K) and above (at 6~K) $T_{\rm c}$. Practically, no difference between the two data sets is observable. The ZF data can be described using a Kubo-Toyabe relaxation function~\cite{Kubo} multiplied by an exponential decay function,
\begin{multline}
A(t)= A(0)\left\{\frac{1}{3}+\frac{2}{3}\left(1-a^2t^2\right){\exp}\left(-\frac{a^2t^2}{2}\right)\right\} \\
{\exp}(-\Lambda t),
 \label{eq:KT_ZFequation}
\end{multline}
where $A(0)$ is the initial asymmetry, and $a$ and $\Lambda$ are muon spin relaxation rates. In the fit, $a$ was kept as a common parameter for both set of data. The fits yield, $a = 0.100(2)$~$\mu$s$^{-1}$, $\Lambda(2{\rm K})=0.025(2)$~$\mu$s$^{-1}$, and $\Lambda(6{\rm K})=0.026(2)$~$\mu$s$^{-1}$. The value of $a$ extracted from the fits reflects the presence of random local fields arising from the nuclear moments within Bi$_4$O$_4$S$_3$. The nearly equal values of $\Lambda$ are consistent with the presence of diluted and randomly oriented electronic moments probably arising from impurities. The ZF-$\mu$SR data completely rule out the presence of any magnetic anomaly in the superconducting state of Bi$_4$O$_4$S$_3$ as for instance the appearance of spontaneous magnetic fields associated with a time-reversal-symmetry breaking pairing state.

\begin{figure}[htb]
\includegraphics[width=1.0\linewidth]{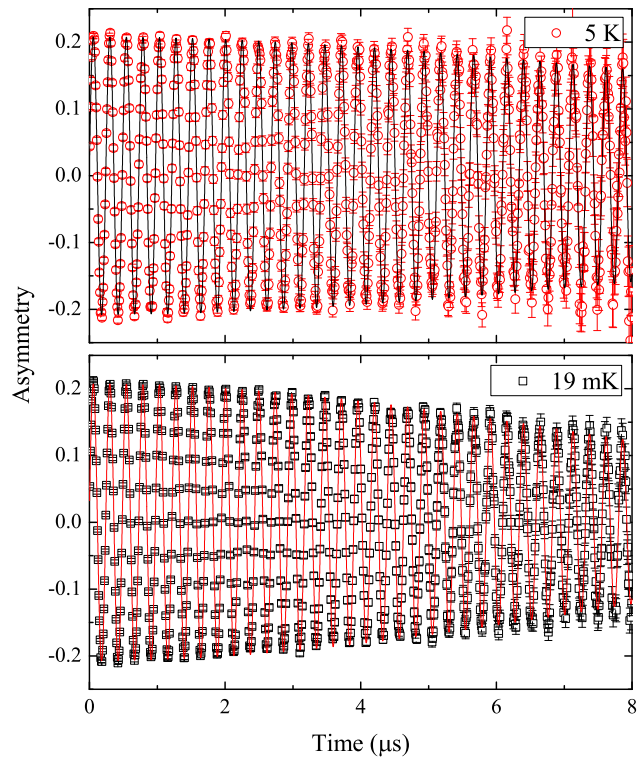}
\caption{(Color online) TF-$\mu$SR time spectra of Bi$_4$O$_4$S$_3$ collected at 19~mK and 5~K in a transverse field of 300~Oe. The solid lines are the fits to the data using the Eq.~\ref{Depolarization_Fit}, described in the text.}
 \label{fig:BiOS_AsyTF}
\end{figure}

Figure~\ref{fig:BiOS_AsyTF} shows the TF-$\mu$SR time spectra collected at 19~mK and 5~K in a transverse field of 300~Oe. At 5~K, the sample is in the normal state and the local field probed by the muons is the applied field slightly broadened by the nuclear moments contribution, which is responsible for the weak relaxation of the precession signal. By contrast, the data collected at 19~mK shows a more pronounced damping due to the inhomogeneous field distribution generated by the formation of a vortex lattice in Bi$_4$O$_4$S$_3$. The TF-$\mu$SR time spectra were analyzed using Gaussian damped spin precession signal along with an exponential relaxation to take into account the corresponding component observed in the ZF spectra.
\begin{multline}
\label{Depolarization_Fit}
A^{TF}(t)=A(0)\exp\left(-\sigma^{2}t^{2}\right/2)\cos\left(\gamma_\mu \left\langle B\right\rangle t +\phi\right){\exp}(-\Lambda_1 t) \\
+A_{\rm bg}(0)\cos\left(\gamma_\mu B_{\rm bg}t +\phi\right),
\end{multline}
where $A(0)$ and $A_{\rm bg}$(0) are the initial asymmetries of the sample and background signals, $\gamma_{\mu}/2\pi=13.55$~kHz/G is the muon gyromagnetic ratio~\cite{Sonier}, $\left\langle B\right\rangle$ and $B_{\rm bg}$ are the internal and background magnetic fields, $\phi$ is the initial phase of the muon precession signal, and $\sigma$ and $\Lambda_1$ are the Gaussian and exponential muon spin relaxation rates, respectively. Since the ZF data indicate that the exponential relaxation rate is temperature independent, we use $\Lambda_1$ as a global common parameter in the TF-$\mu$SR data fitting and obtain a value of 0.038(3)~$\mu$s$^{-1}$. The background signal mainly comes from the silver sample holder where the relaxation  rate is
negligible.

\begin{figure}[htb]
\includegraphics[width=1.0\linewidth]{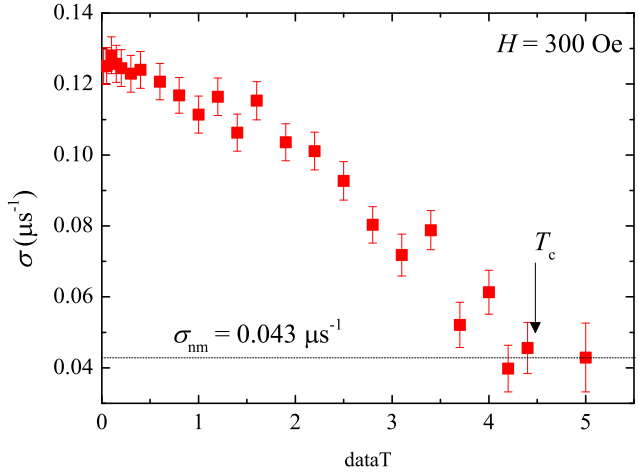}
\caption{(Color online) The temperature dependence of $\sigma$ of Bi$_4$O$_4$S$_3$ for an applied field of 300~Oe. The dotted line parallel to the temperature axis shows the temperature independent value of $\sigma_{\rm nm}$, the nuclear magnetic dipolar contribution to $\sigma$.}
 \label{fig:BiOS_Sigma_v1}
\end{figure}

Figure~\ref{fig:BiOS_Sigma_v1} shows the temperature dependence of $\sigma$ for an applied field of 300~Oe. The Gaussian muon spin relaxation rate $\sigma$ is proportional to the second moment of the internal field distribution sensed by the muons and can be expressed as $\sigma=\left(\sigma^{2}_{\rm sc} + \sigma^{2}_{\rm nm}\right)^{\frac{1}{2}}$, where $\sigma_{\rm sc}$ is the contribution due to the inhomogeneous field distribution in the vortex state and $\sigma_{\rm nm}=0.043(1)$~$\mu$s$^{-1}$ is the nuclear magnetic dipolar contribution which is assumed to be temperature independent.

\begin{figure}[htb]
\includegraphics[width=1.0\linewidth]{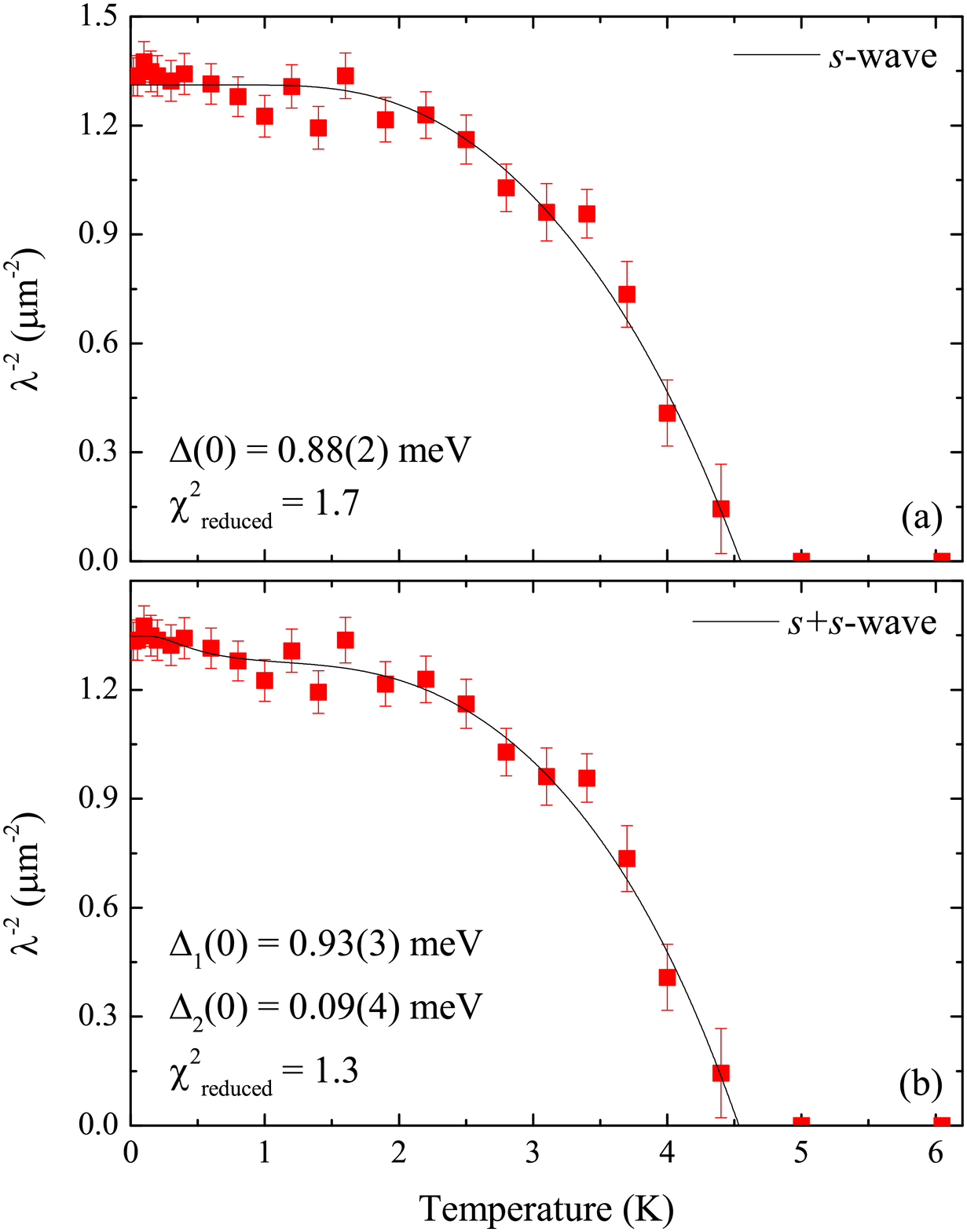}
\caption{(Color online) Temperature dependence of $\lambda^{-2}$ for Bi$_4$O$_4$S$_3$. The curves are fits to the data using a (a) single- and (b) a two-gap $s$-wave model.}
 \label{fig:BiOS_Lambda_v1}
\end{figure}

For a superconductor with hexagonal vortex lattice, $\sigma_{\rm sc}$ is related to the penetration depth, $\lambda$ by the Brandt equation~\cite{Brandt2}, which is based on a Ginzburg-Landau treatment of the vortex state,
\begin{equation}
\sigma_{\rm sc}(b)[\mu{\rm s}^{-1}]=4.854\times10^4 (1 - b) [1 + 1.21(1 -\sqrt{b})^3]\lambda^{-2}[{\rm nm}^{-2}],
 \label{eq:Brandt_equation1}
\end{equation}
Here $b=\left\langle B\right\rangle/B_{\rm c2}$ is a reduced magnetic field and $\left\langle B\right\rangle$ is the first moment of the field distribution, which in our case is very close to the applied field. Equation~\ref{eq:Brandt_equation1} was used to calculate $\lambda^{-2}(T)$, which is proportional to the superfluid density. Figure~\ref{fig:BiOS_Lambda_v1} shows the temperature dependence of $\lambda^{-2}$ for Bi$_4$O$_4$S$_3$. The fits to the $\lambda^{-2}(T)$ data were then made with a single- and a two-gap $s$-wave model using the following functional form~\cite{Carrington, Padamsee}:

\begin{equation}
\label{two_gap}
\frac{\lambda^{-2}\left(T\right)}{\lambda^{-2}\left(0\right)}=\omega\frac{\lambda^{-2}\left(T, \Delta_{1}(0)\right)}{\lambda^{-2}\left(0,\Delta_{1}(0)\right)}+(1-\omega)\frac{\lambda^{-2}\left(T, \Delta_{2}(0)\right)}{\lambda^{-2}\left(0,\Delta_{2}(0)\right)},
\end{equation}
where $\lambda\left(0\right)$ is the value of the penetration depth at $T=0$~K, $\Delta_{\rm i}(0)$ is the value of the $i$-th ($i=1$ or 2) superconducting gap at $T=0$~K and $\omega$ is the weighting factor of the first gap.

Each component of Eq.~\ref{two_gap} can be expressed within the local London approximation ($\lambda \gg \xi$)~\cite{Tinkham,Prozorov} as

\begin{equation}
\frac{\lambda^{-2}\left(T, \Delta_{\rm i}(0)\right)}{\lambda^{-2}\left(0, \Delta_{\rm i}(0)\right)}=1+2\int^{\infty}_{\Delta_{\rm i}(0)}\left(\frac{\partial f}{\partial E}\right)\frac{ EdE}{\sqrt{E^2-\Delta_{\rm i}\left(T\right)^2}},
\end{equation}
where $f=\left[1+\exp\left(E/k_{\rm B}T\right)\right]^{-1}$ is the Fermi function, and $\Delta_{\rm i}\left(T\right)=\Delta_{\rm i}(0)\delta\left(T/T_{\rm c}\right)$. The temperature dependence of the gap is approximated by the expression $\delta\left(T/T_{\rm c}\right)=\tanh\left\{1.82\left[1.018\left(T_{\rm c}/T-1\right)\right]^{0.51}\right\}$~\cite{Carrington}.

The curves shown in Fig.~\ref{fig:BiOS_Lambda_v1}~(a)~and~(b) are the fits to the $\lambda^{-2}(T)$ data using the two models. Table~\ref{table_of_gapratios} summarizes the corresponding fitted parameters. The two \textit{s}-wave gap model gives a lower $\chi_{\rm reduced}^2$ value than a single gap model thus suggesting that Bi$_4$O$_4$S$_3$ is a multiple gap superconductor and making a connection between BiS$_2$-based layered superconductors and the layered Fe-based superconductors~\cite{Khasanov,Biswas2,Biswas3,Paglione}. Other prominent examples of superconductivity are represented by MgB$_2$, NbSe$_2$, and Lu$_2$Fe$_3$Si$_5$~\cite{Choi,Yokoya,Nakajima,Biswas4}. However, a single $s$-wave gap can not be ruled out completely as claimed in ref.~\onlinecite{Shruti}.

From our data, we obtain $\lambda(0)=861(17)$ nm. Remarkable is that the value of lambda at $T=0$~K is one of the highest among all known bulk superconductors. This reflects a very low superfluid carrier density $n_s$ of $\approx 3 \times 10^{21}$ cm$^{-3}$. It should be noted that the value of $\lambda(0)$ is independent on details of analysis such as the correction for the field dependence of $\sigma$ and it is not strongly dependent on the assumption of a vortex lattice of hexagonal symmetry. The assumption of a regular vortex structure is generally valid in the FC case if parameters such as the external field are chosen in a reasonable range and if the system is not extremely anisotropic. Having hexagonal or rectangular symmetry makes almost no difference in the value of the penetration depth and its temperature dependence. For instance, for a square arrangement the pre-factor of Eq.~\ref{eq:Brandt_equation1} would change only by 2 \%. The effect of vortex disorder is more difficult to quantify exactly. Disorder would give an additional contribution to  $\sigma$  so that the estimated value for  $n_s$ is actually an upper limit. Since the contribution of disorder adds quadratically to that of the intrinsic inhomogeneity of the vortex state we do not expect a big change of the value of $\lambda$. Even in systems, such as $R$(Fe$_{1-x}$Co${x}$)$_{2}$As$_{2}$ where disorder may be important, measurements on the same system in the vortex state by TF-$\mu$SR \cite{Williams} (sensitive to disorder)  and in the Meissner state by low energy $\mu$SR \cite{Ofer} (insensitive to this effect) gave very close results for $\lambda$ and its temperature dependence.

\begin{table}
\caption{Fitted parameters of the fits to the $\lambda^{-2}(T)$ data of Bi$_4$O$_4$S$_3$ using different models.}
\label{table_of_gapratios}
\begin{center}
\begin{tabular}[t]{llll}\hline\hline
Model & Gap value & Gap ratio & $\chi_{\rm reduced}^2$ \\
{} & $\Delta(0)$ (meV) & $\Delta(0)/k_{\rm B}T_{\rm c}$ & {} \\\hline
\textit{s}-wave & 0.88(2) & 2.25(5) & 1.7\\
\textit{s}+\textit{s}-wave~~~~& 0.93(3), 0.09(4)~~~~& 2.38(7), 0.22(9)~~~~& 1.3\\
{} & with $\omega=0.94(1)$ & {} & {}\\\hline\hline
\end{tabular}
\par\medskip\footnotesize
\end{center}
\end{table}
%


In summary, resistivity and $\mu$SR measurements have been performed on superconducting Bi$_4$O$_4$S$_3$. Temperature dependence of the upper critical fields and also the absolute value, $H_{\rm c2}(0)=15.3(4)$~kOe were determined from the resistivity measurements. The $H_{\rm c2}(T)$ data shows a positive curvature close to $T_{\rm c}$, which is a sign of multi-gap superconductivity in Bi$_4$O$_4$S$_3$. ZF-$\mu$SR data show no sign of magnetic anomaly in the superconducting ground state of Bi$_4$O$_4$S$_3$. The temperature dependence of $\lambda^{-2}$ is compatible with a two-gap $s$-wave model with the zero-temperature gap values, $\Delta_{1}(0)=0.93(3)$, $\Delta_{2}(0)=0.09(4)$~meV and penetration depth, $\lambda(0)=861(17)$~nm. The presence of two superconducting gaps in Bi$_4$O$_4$S$_3$ is predicted by theoretical works such as first principles band structure calculations suggesting two band crossing the Fermi surface and the RPA analysis of a two-orbital model for the BiS$_2$-based superconductors~\cite{Usui,Martins}. Hall, Seebeck coefficients and magnetoresistance measurements also further support our experimental finding~\cite{Li,Tan}. It is worth mentioning that even though our data are more compatible with a two-gap model, a single-gap model can not be ruled out completely. Our results show that Bi$_4$O$_4$S$_3$ is a bulk superconductor with a fully developed node-less superconducting energy gap and extremely low superfluid density compared to other layered superconductors.

The $\mu$SR experiments were performed at the Swiss  Muon Source (S$\mu$S), Paul Scherrer Institute (PSI, Switzerland). Work at Brookhaven is supported by the Center for Emergent Superconductivity, an Energy Frontier Research Center funded by the DOE Office for Basic Energy Science (H.L. and C.P.). P.K.B. would like to acknowledge M. Medarde and R. Sibille for their assistance in the transport measurements.

\end{document}